\documentclass[fleqn,usenatbib]{mnras}

\usepackage{newtxtext,newtxmath}

\usepackage[T1]{fontenc}

\DeclareRobustCommand{\VAN}[3]{#2}
\let\VANthebibliography\thebibliography
\def\thebibliography{\DeclareRobustCommand{\VAN}[3]{##3}\VANthebibliography}

\usepackage{graphicx}	
\usepackage{amsmath}	
\usepackage{amssymb}	
\usepackage{threeparttable} 
\usepackage{caption}    
\usepackage{subcaption} 

\title[Ripples on a Big Screen]{Hubble's Variable Nebula I: Ripples on a Big Screen}

\author[J. F. Lightfoot]{
John Lightfoot,$^{1}$\thanks{E-mail: johnlightfoot1957@gmail.com (JFL)}
Aleks Scholz,$^{2}$
\\
$^{1}$3 East Newington Place, Edinburgh EH9~1QP, UK\\
$^{2}$SUPA, School of Physics \& Astronomy, University of St~Andrews, North Haugh, St~Andrews\\ KY16~9SS, UK\\
}

\date{Accepted 2025 April 23. Received 2025 April 23; in original form 2024 December 12}

\pubyear{2025}

\begin{document}
\label{firstpage}
\pagerange{\pageref{firstpage}--\pageref{lastpage}}
\maketitle

\begin{abstract}
NGC~2261 is a reflection nebula illuminated by the young star R~Monocerotis. Objects moving near the star occasionally cast shadows on the nebula, giving rise to its alternative name: Hubble's Variable Nebula. 

For 7 years since Spring 2017 robotic telescopes have been used to compile a roughly twice-weekly record of changes in the object. The results, over 1000 images at separate epochs, have been compiled into a movie. This shows that, as well as the large scale but infrequent variability for which it is famous, the nebula is continually traversed by low level `ripples' of light and dark. These record changes in the light output from R~Mon and analysis of their progress indicates that the reflecting material takes the form of a thin ($< 3\times10^{16}$~cm) screen whose shape resembles a half paraboloid, rooted at the star and bowed towards us.

The brightness of the screen in {\em Herschel} far-IR maps indicates a density n$_H > 1.7\times10^5$~cm$^{-3}$ and CO observations show the material is moving towards us at a few km~s$^{-1}$ relative to the rest cloud, consistent it with being a dense shell of material displaced by R~Mon's outflow.

The results demonstrate the value of studying such objects in the time domain, and are a glimpse of what will be achieved by instruments like the Zwicky Transient Facility and Vera Rubin Observatory.
\end{abstract}

\begin{keywords}
stars: variables: Herbig Ae/Be -- (stars:) circumstellar matter -- protoplanetary discs
\end{keywords}



\section{Introduction}

NGC~2261 is a reflection nebula illuminated by the Herbig Ae/Be star R~Monocerotis \citep{1960ApJS....4..337H}. The star is obscured at optical wavelengths \citep{1968ApJ...152..439H}, a source of strong infra-red radiation \citep{1966ApJ...143.1010M}, linked to optical and molecular outflows aligned with Herbig-Haro 39 \citep{1982AJ.....87.1223J}, and associated with a molecular cloud \citep{canto1981carbon}. The outflows give dynamical ages ranging from 5900 to $2.8\times10^5$~yr \citep{Sandell_2020}. The common inference has been that R~Mon is a young star still interacting with the cloud from which it formed.

The nebula is associated with a small dark cloud lying $1^{\circ}$ away from that containing the young cluster NGC 2264, whose distance has been derived from {\em Gaia} parallax measurements to be $719\pm16$~pc \citep{Ma_z_Apell_niz_2019}. That the apparent association of the clouds is real is reinforced by their similar CO radial velocities \citep{1978ApJ...226..839C}.

NGC~2261 is famous for its variability. Changes in the fan were first noticed in 1915 by John Mellish, astronomer assistant at Yerkes Observatory \citep{2000AAS...197.0105W}, and confirmed by the young Edwin Hubble \citep{1916ApJ....44..190H}, since when the object has also been known as "Hubble's Variable Nebula".

The variability was studied by Carl Lampland at Lowell Observatory who amassed 940 photographs of the object between 1916 and 1951. In 1967 Richard C. Hall at Arizona State University compiled a selection of them into a movie \footnote{To find the movie, search the Internet for "Hubble's Variable Nebula NGC 2261 C.O. Lampland's Photographic Record".}.

NGC~2261 belongs to a small class of variable nebulae. Other members include NGC~1555 illuminated by young star T~Tauri, discovered by John Russell Hind in 1852 then observed to fade over the next few years \citep{1936PASP...48..318L}, and which Zwicky Transient Facility (ZTF) \citep{2019PASP..131a8002B} observations show still varying today. A Southern representative is NGC~6729 near R~CrA, discovered to be variable by \citet{1920HelOB..20..182K} in 1913 and still performing in 1984 \citep{1987PASP...99...91G}. 

The speed and scale of NGC~2261's variations led \citet{bellingham} to suggest they are shadows cast onto the illuminated fan by objects moving near the star. If this is the case, observing the shadows is one step away from seeing the shadowing objects themselves. \citet[][]{10.1093/mnras/239.2.665} tried to do this using the Hall movie but it was clear that regular observation with digital cameras was needed to improve the data. The advent of robotic telescopes made this feasible and the monitoring campaign reported here was begun in March 2017.

\section{Observations}
The aim of the observations was to catch variations in NGC~2261 and sample them finely enough to trace their development accurately. This entailed a campaign lasting several years with a sampling period of once per week or better, as movement of dark shapes near R~Mon at $1/4''$ per day had been reported by Lampland. 

Most of the observations reported here were requested from the 1-m network of the Las Cumbres Observatory Global Telescope (LCOGT) \citep{2013PASP..125.1031B} using time from several iterations of program `Time Domain Observations of Young Stars' (PI: A. Scholz). Others were obtained from the LCOGT data archive, the Liverpool Telescope (LT) archive \citep{2004SPIE.5489..679S}, and the Faulkes Telescope Project \citep{lewis2009faulkes}. Arizona astronomer Tom Polakis plugged some gaps in the sequence when the Covid pandemic knocked out the robots. 

\begin{figure}
	    \includegraphics[width=\columnwidth]{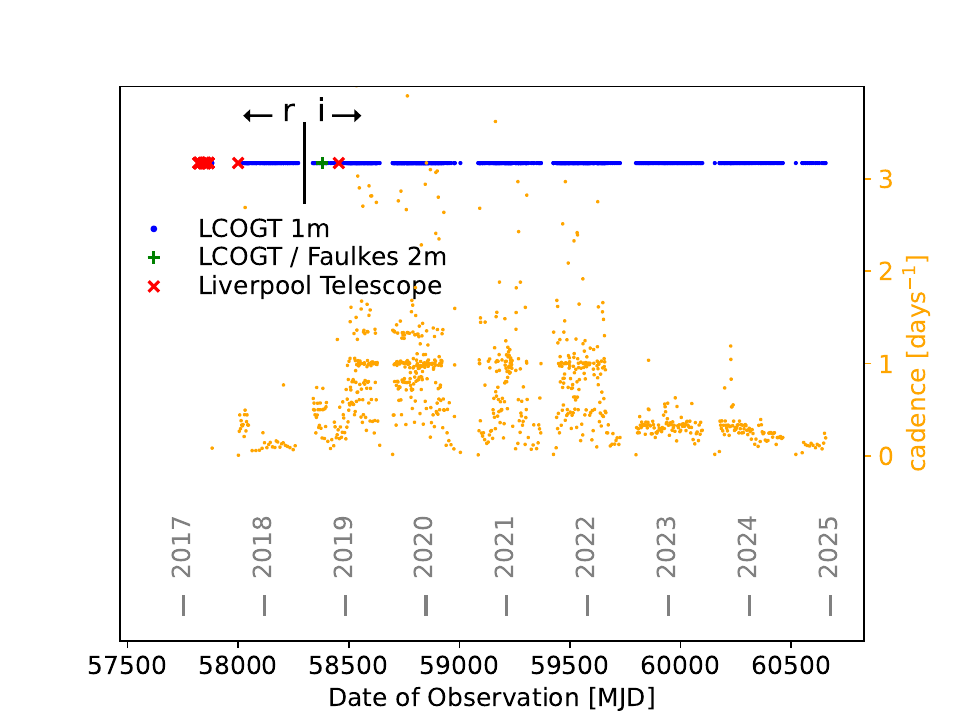}
        \caption{The origin, epoch and cadence of observations described in this paper. The vertical bar at MJD~58300 shows where the filter used switched from r$^\prime$ to i$^\prime$.}
        \label{fig:observations}
\end{figure}

Observation details are summarised in Table~\ref{tab:observations} and Fig.~\ref{fig:observations}. The dashed `line' along the top of Fig.~\ref{fig:observations} is the unresolved blend of more than 1000 individual observation points, running from 2017 to 2024 with gaps each summer as the Sun moved near Monoceros. The plotted `cadence' is the inverse of the time between adjacent observations, so that higher numbers mean more frequent observation. The demand cadence varied between 0.25 (observation every 4 days) and 1 (observation daily) but scheduling vagaries added noise. A scattering of points with higher cadence were either intentional repeats of a subpar image, automatic repeats generated for some reason by the robot scheduler, or simply mistakes. 
\begin{table}
	\centering
	\begin{threeparttable}
	\begin{tabular}{lcccr}
		\hline
		Facility & Camera & Filters & Exp(s) & \# \\
		\hline
      	LCOGT 1m & Sinistro & i$^\prime$ & 250 & 964\\
		LCOGT 1m & Sinistro & r$^\prime$ & 250 & 168\\
		LT 2m & IO:O & r$^\prime$ & $4\times17$ & 8\\
		Faulkes 2m & Spectral & i$^\prime$ & 50 & 4\\
		Polakis 12.5" & SBIG  & clear & $4\times300$ & 2\\
		\hline
	\end{tabular}
	\end{threeparttable}
    \caption{Summary of observations.}
    \label{tab:observations}
\end{table}

\section{Data Reduction}
The data were reduced as follows:
\begin{enumerate}
    \item Manually remove cosmetically bad images. 
    \item Calibrate image astrometry and photometry with {\em SExtractor} \citep{SEXTRACTOR1996} and {\em Scamp} \citep[][]{SCAMP2006}, using reference images from the Sloan Digital Sky Survey \citep{2000AJ....120.1579Y}. On rare occasions {\em Scamp} failed to match image stars with the reference and the fix was to re-calibrate the image WCS with {\em Astrometry.net} \citep{astrometry.net}.  
    \item Resample images to a common grid with {\em Swarp} \citep[][]{2002ASPC..281..228B}.
    \item Combine the image sequence into a cube with axes [epoch:RA:Dec]. The epoch axis is regular with a spacing of 1 day but no interpolation in time between images was performed: the plane at a nominal date contains the image with the nearest actual date, unless none could be found closer than 10 days when the frame was left blank. 10 days is a compromise chosen to paper over unavoidable discontinuities and let the movie flow, without becoming too inaccurate. The sampling timescale of between a few and 10 days is adequate for following the action higher up the nebula, but things are seen to move quickly ($\sim1''/$~day) within $10''$ of R~Mon and there the changes were not sampled so well. 
    \item Convert the image sequence cubes to movies by plotting their planes and combining these with {\em ffmpeg} \citep{tomar2006converting}. Each movie frame is tagged with the Modified Julian Day (MJD) and filter of the observation used.
\end{enumerate}

\section{The Variability}
\begin{figure}
	    \includegraphics[width=\columnwidth]{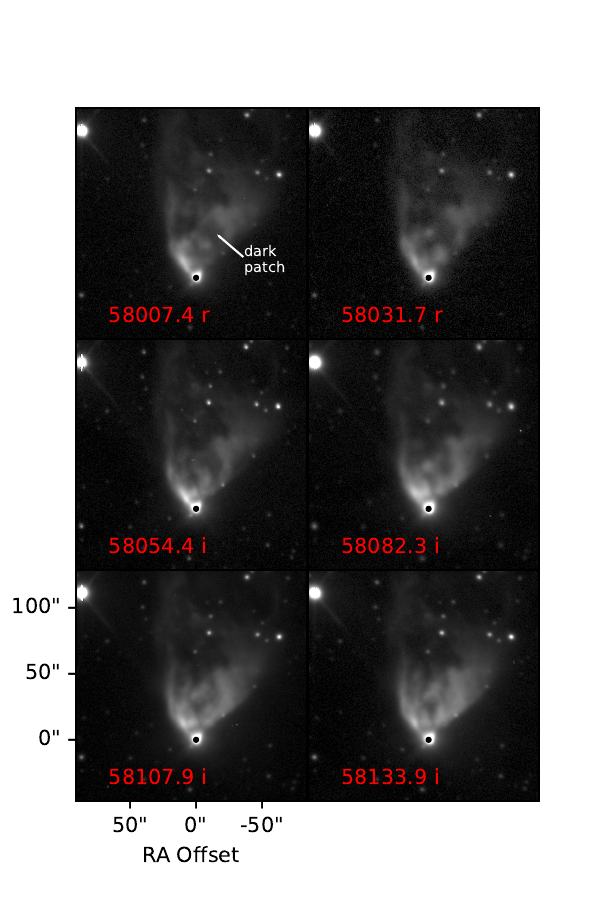}
        \caption{Frame sequence showing strong nebula variation in SDSS r$^\prime$ and i$^\prime$ in late 2017 / early 2018. A black dot marks R~Mon and the initial position of the dark patch is indicated. A movie of the sequence is in the supplementary material online.}
        \label{fig:straight_movie}
\end{figure}

The nebula's signature variability consists of dark patches that on occasion stretch across and move up the reflection fan. The end of one such event was caught at the start of the campaign and is shown as a frame sequence in Fig.~\ref{fig:straight_movie}. The moving patches are difficult to see against the nebula's static pattern in isolated frames so the initial position of the dark patch has been marked. 

\begin{figure}
	    \includegraphics[width=\columnwidth]{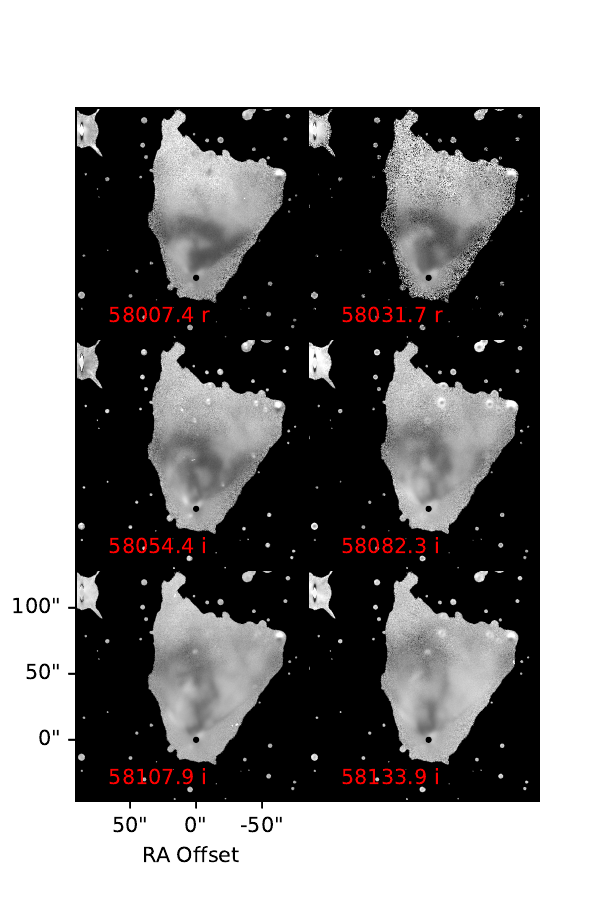}
        \caption{The same frame sequence as Fig.~\ref{fig:straight_movie} where each pixel has been divided by its median over time. Areas with low median value are blank. A movie of the sequence is in the supplementary material online.}
        \label{fig:normalised_movie}
\end{figure}

To make the variations easier to see, Fig.~\ref{fig:normalised_movie} shows a modified view of the event where each frame has been divided by the nebula median over time; this removes the static pattern so that the main patches are clear and many subtle details are revealed.

By far the best way to appreciate the data is to watch the movies, which are available as supplementary material online. The human eye can see and follow details that are hard to appreciate otherwise. 

\begin{figure}
	    \includegraphics[width=\columnwidth]{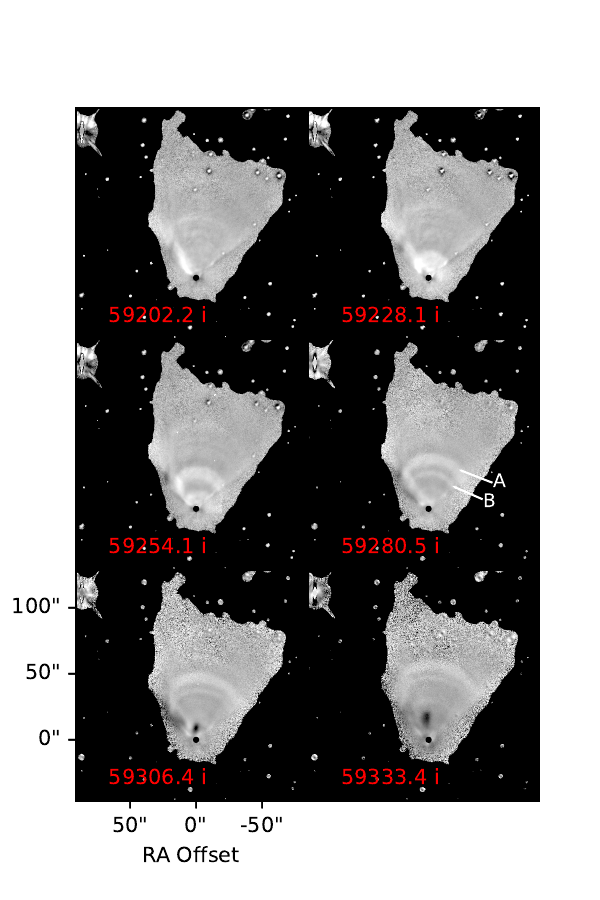}
        \caption{Frame sequence of ripples A and B analysed in Section~\ref{sec:big-screen} to derive the shape of the `screen'.}
        \label{fig:ripple_sequence1}
\end{figure}

The variation event shown in Figs.~\ref{fig:straight_movie} and \ref{fig:normalised_movie} occurred right at the start of the observing campaign and exhibits some typical behaviours already noted by Lampland and Hall: patches generally start low on the W limb, sweep across and up the fan to the E limb, becoming confined to the E side as they move up and eventually out of the nebula. This event had ended by MJD~58340, then dark patches returned on a smaller scale before leaving again by MJD~58640. Since that time the nebula has been quiet except for some dark murmurings close to R~Mon, notably a dark blob swinging across near MJD~59306 that is visible in the last two panes of Fig.~\ref{fig:ripple_sequence1}. 

The lack of strong activity in recent years is disappointing but does let more subtle variations be seen. Most notably ripples of light and dark constantly traverse the fan which are drowned out during more vigorous passages. These have not been noted before and what they tell us about the distribution of the nebula's reflecting material will form the main focus of this paper. The changes in R~Mon's brightness that cause the ripples is also of interest and will be the subject of future work.

\section{The Big Screen}
\label{sec:big-screen}
If the variability dark patches are indeed shadows then the promise of their study is that we can `observe' the objects responsible, which lie closer to R~Mon than can be seen directly. To do this we need to know the delay of the light echo from the fan relative to the direct ray from the star. The delay can range up to several months and the precise value depends on where the reflecting material is located. 

First, how is the material distributed: does it fill a cloud, concentrate in a group of blobs, or spread across a smooth surface? The behaviour of the newfound ripples answers this question directly. Their lack of directional variation implies they are caused by changes in the overall brightness of R~Mon, and they never cross each other, which points to the scattering dust being arranged in a single layer - multiple surfaces would give rise to families of ripples moving at different apparent speeds. We shall refer to the single layer as the `screen'.

The regular arc and movement of the ripples indicate that the screen presents a smooth face on a large scale, though it is also variable in density as shown by the veils, blobs and filaments in HST images of the object.

The shape of the screen can be derived from the movement of the ripples. Two were studied in detail, marked A and B in Fig.~\ref{fig:ripple_sequence1}. These were chosen because their high contrast and apparently sharp edges made them easy to measure, which was done by eye, moving and clicking the cursor along the ripple edges visible in each image of the sequence. 

\begin{figure}
    \includegraphics[width=\columnwidth]{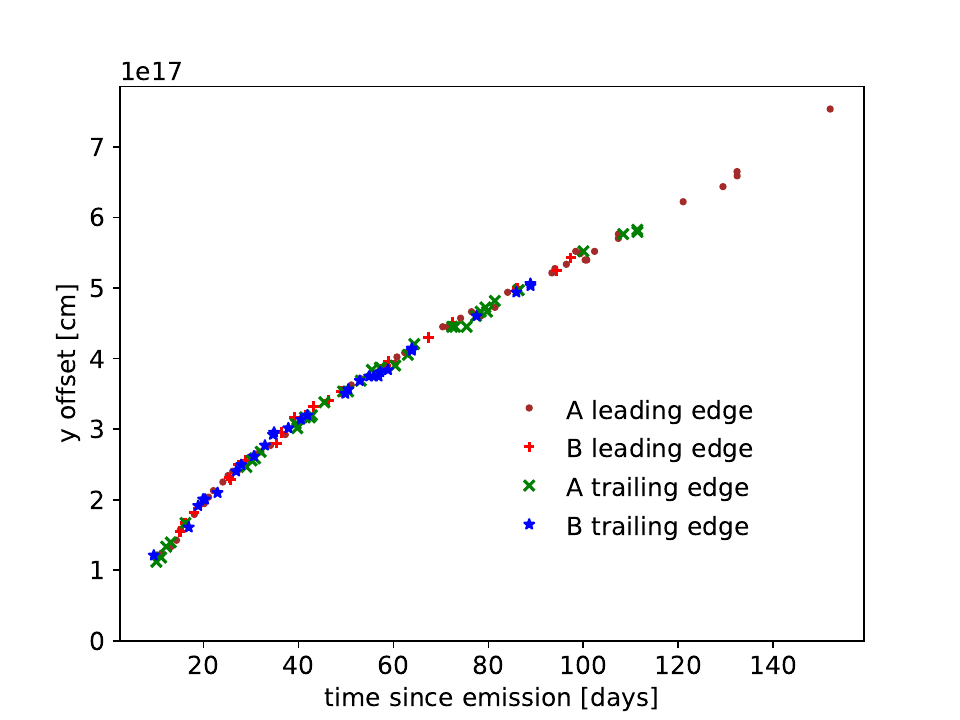}
    \caption{Progression of the edges of 2 bright ripples on a line N from R~Mon. The y-axis assumes a distance of 716~pc. The x-axis is time since the ripple edge left the star.}
    \label{fig:bright_ripples}
\end{figure}

Fig.~\ref{fig:bright_ripples} plots the ripple progress N from R~Mon versus time, with lengths calculated assuming a distance to the nebula of 719~pc. The four different colours and symbols in Fig.~\ref{fig:bright_ripples} represent measurements of the leading and trailing edges of the 2 ripples: time offsets have been applied to make the curves lie on top of each other. The four tracks are identical within the limits of measurement, the scatter giving an idea of the error. It is not possible to measure the ripples close to R~Mon so the origin of the time axis is a guess with a possible error of 3 days or so.

\begin{figure}
    \includegraphics[width=\columnwidth]{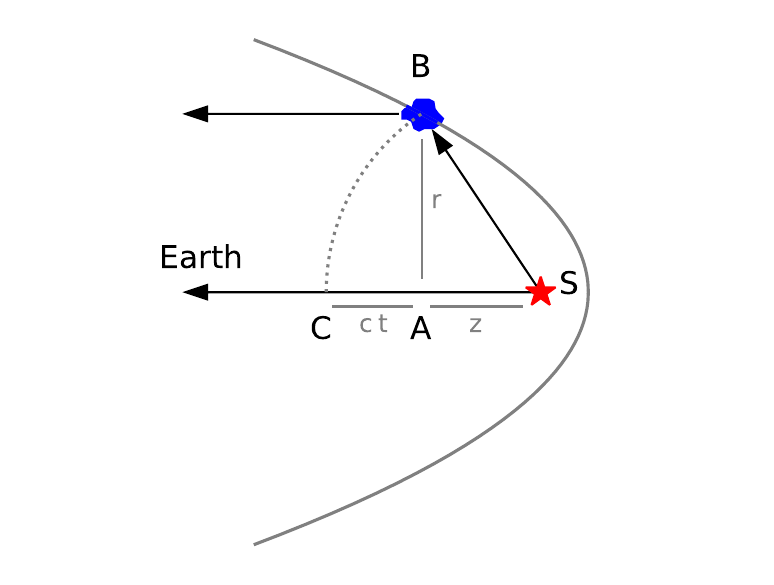}
    \caption{The geometry of a light echo. S is the star, B the scatter point, the curved solid line the echo ellipse.}
    \label{fig:geometry}
\end{figure}

How does ripple progress relate to the screen shape? A characteristic of any ellipse is that the sum of the distances from any point on its locus to the two foci is constant. Therefore, the echo seen on Earth at a particular time after pulse emission must have scattered from somewhere on a particular ellipse with R~Mon at one focus and Earth at the other. \citet{couderc} derived an expression for the position of the scatter point on the ellipse. For the geometry shown in Fig.~\ref{fig:geometry}: where S is the star, B the scatter point, the dotted line the wavefront of a light pulse from star, and line SA points to the observer: the $z$ offset of the clump is given by:
\begin{equation}
    \label{eq:couderc}
    z = \frac{r^2}{2ct} - \frac{ct}{2}
\end{equation}
where $r$ is the apparent offset of the clump, $t$ the delay between arrival of the direct ray and the echo, and $c$ the speed of light. The difference in path length of A->Earth and B->Earth is neglected as being small: $\sim r \times r_{ang}$, where $r_{ang}$ is the clump angular offset. 

\begin{figure}
    \includegraphics[width=\columnwidth]{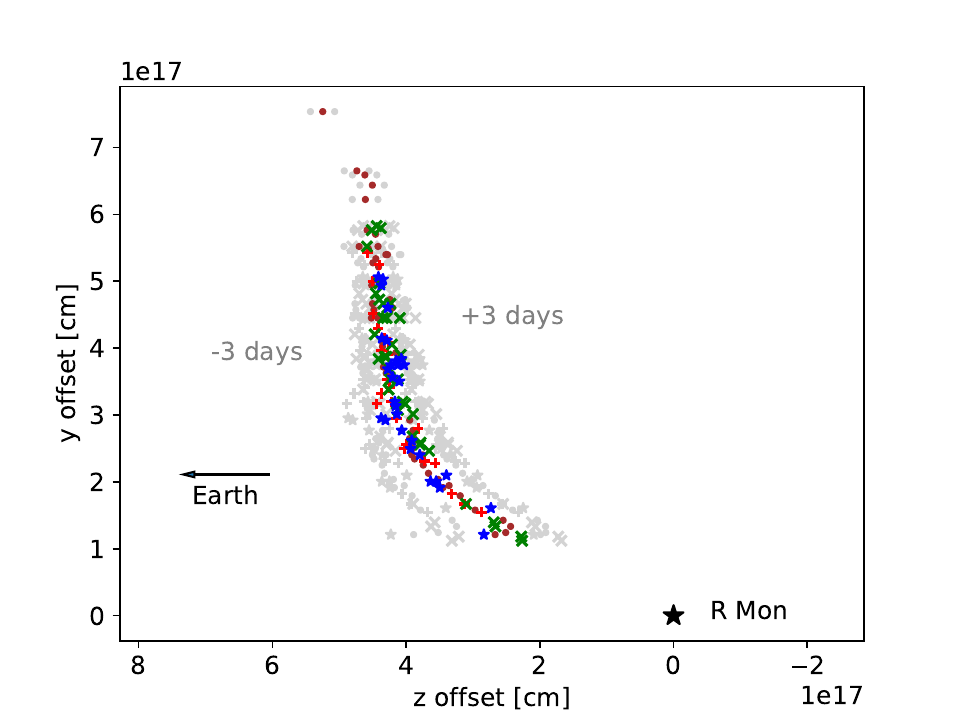}
    \caption{The shape of the reflecting surface on the line N from R Mon. The gray points to left and right are results for the `time since emission' axis in Fig.~\ref{fig:bright_ripples} shifted by $\pm3$ days.}
    \label{fig:meridian_shape}
\end{figure}

Using Eqn.~\ref{eq:couderc}, $z$ offsets corresponding to the $y$ values in Fig.~\ref{fig:bright_ripples} were calculated and plot in Fig.~\ref{fig:meridian_shape}. The effect of the 3 day uncertainty in the origin of the time axis is shown by the gray points, which show the $z$ offsets for time origin shifted by $\pm3$ days. Shifts greater than 3 days lead to curves that look unphysical as the bottom end of the reflecting screen no longer points toward R~Mon. 

A maximum limit to the screen thickness can also be obtained. Differentiating Eqn.~\ref{eq:couderc} and inserting values for $r$ and $t$ appropriate to the top and bottom of the curve in Fig.~\ref{fig:bright_ripples} gives values for $dz/dt$ of $-0.8\times10^{11}$ and $-3.0\times10^{11}$~cm/s respectively. If the screen has uniform thickness and we assume that the leading/trailing edge echos come from the near/far side of the layer then the variation of $dz/dt$ with $r$ will change the shape of the curve in Fig.~\ref{fig:bright_ripples} as the thickness increases. This means the leading and trailing edge data could not be shifted to lie on top of each other, there would be a split at one end or the other. No split is visible in Fig.~\ref{fig:bright_ripples} to a resolution of roughly 3 days, corresponding to a maximum screen thickness of $3\times10^{16}$~cm.

\begin{figure}
    \includegraphics[width=\columnwidth]{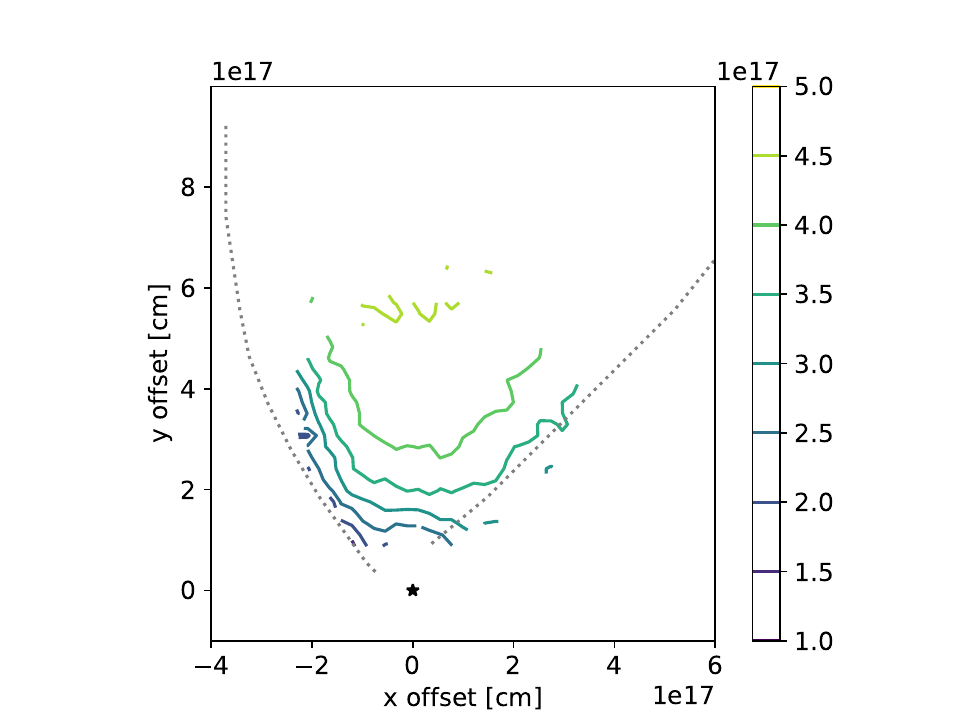}
    \caption{Contour plot showing the $z$ offset [cm] relative to R~Mon of the screen in NGC~2261. Positive $z$ is towards Earth. The fan outline is shown dotted.}
    \label{fig:surface}
\end{figure}

Extending the analysis to 2 dimensions, measurements of the ripples across the fan led to the result shown in Fig.~\ref{fig:surface}. The $z$ contours clearly show a screen that is rooted at R~Mon and tilted towards the Earth to the N; it curves into the line of sight at the E limb and less so towards the W limb. The curve of the screen subtends roughly $15^{\circ}$ at the star and any object moving within that arc will cast a visible shadow.

\section{Other Observations}
How does the `screen' fit with other observations?

At visible wavelengths the nebula is fan shaped with R~Mon at the apex, though there is an E-W asymmetry, with the E limb soft and curved like a cavity boundary while the W is sharp and geometrical like a shadow across a smooth surface. A deep image \citep{10.1093/mnras/217.1.31} shows the fan edges becoming parallel further N from the star and fading as they approach HH-39 $7'$ away, which presumably marks the termination of the stellar outflow. A windswept cavity expanding N from R~Mon and becoming a hollow tube at distance is consistent with the shape of the screen, which covers the area of the expanding cone with the W limb shadowed and therefore not measured.  

The screen is in accord with the work of \citet{1991MNRAS.249..707M} who modelled the nebula's {\em J}, {\em H} and {\em K} polarization and concluded that the blue-shifted outflow to the N of the star is a parabolic cavity, with the scattering of radiation taking place mainly at the cavity surface.





\subsection{The far-infrared Nebula}
\label{sec:FIR_nebula}
Far-IR maps are deeply revealing of overall cloud structure because thermal emission from dust is ubiquitous and generally optically thin.

NGC~2261 was observed by {\it Herschel}. Continuum maps were made with the PACS \citep{poglitsch} at wavelengths~(HPBW) of $70\micron~(5\farcs6)$, $100\micron~(6\farcs8)$, $160\micron~(11\farcs3)$, and with SPIRE \citep{griffin} at $250\micron~(18\farcs4)$, $350\micron~(25\farcs2)$, and $500\micron~(36\farcs7)$. A {\it Spitzer} map at $24\micron$ of the region near HH-39 shows what may be the upper wall of the NGC 2261 reflection nebula cavity \citep{2007astro.ph..1906A}.

To summarise the analysis below, the {\em Herschel} data are consistent with a dusty screen bounding a cavity N of R~Mon. The visible fan is starlight scattered by the grains while the IR fan is thermal emission from them warmed by light absorbed. The IR fan's bright E edge marks where the cavity wall curves into the line of sight so that it is limb brightened. The screen is denser than its surroundings with $n_H > 1.7\times10^5$~cm$^{-3}$. The counter-fan S of R~Mon is relatively empty.

\begin{figure}
    \includegraphics[width=\columnwidth]{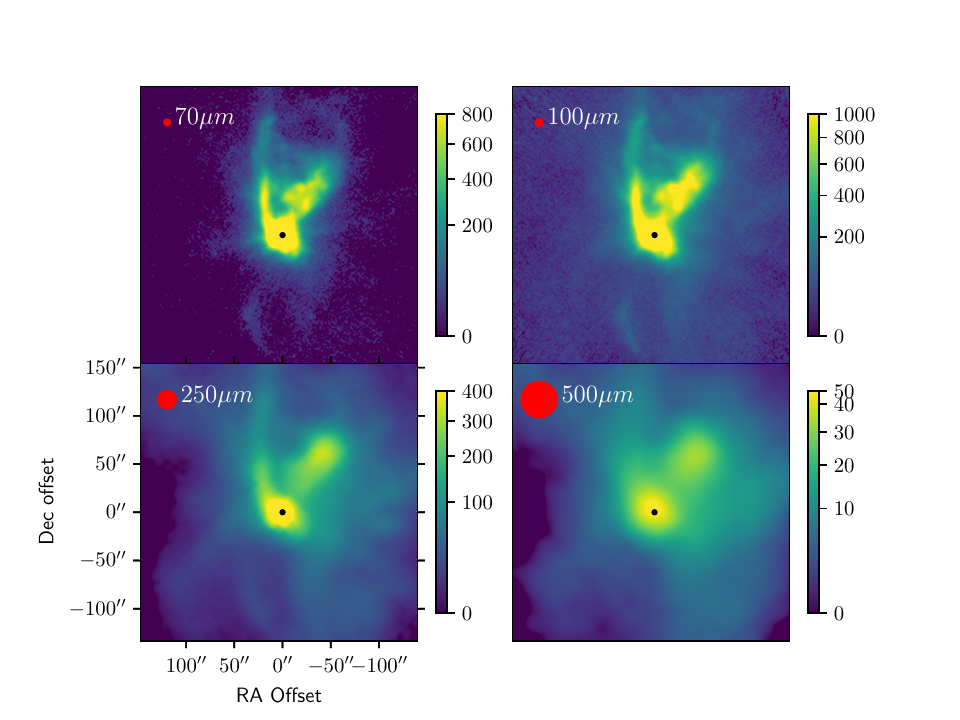}
    \caption{The nebula at selected {\em Herschel} wavelengths [MJy/sr]. The telescope beam is indicated in red and R~Mon is marked by a black spot.}
    \label{fig:farIR}
\end{figure}

In more detail, a selection of the {\it Herschel} maps are presented in Fig.~\ref{fig:farIR}. These are calibrated data products taken directly from the archive. All show a peak at R~Mon, with a bright extension NE from the star along the E limb of the visible fan. At $70\micron$ emission N of the star generally matches the shape of the fan, while S of the star there is only a faint wisp - which coincides with a feature in deep visible-light images. A second peak $80\arcsec$ NW of R~Mon has no obvious visible counterpart and is stronger at long wavelengths. At $500\micron$ the nebula is a bright patch on a broad plateau of emission from cool material.

The strength of far-IR thermal emission depends on the optical properties, column density and temperature of the dust responsible. If the optical properties are known, each wavelength-adjacent pair of maps can be used to estimate the column density as follows:
\begin{enumerate}
    \item Blur the shorter wavelength map to match the resolution of the longer by convolving it with a Gaussian of width:
    \begin{equation}
        FWHM = \sqrt{(FWHM_{long}^2 - FWHM_{short}^2)}
    \end{equation}
    \item Estimate the colour temperature of the emission $T_d$ from the ratio $r$ of the 2 maps:
    \begin{equation}
        r = \frac {\frac{ \int_{short}^{} B_\nu (T_d) Kabs_\nu d\nu} {\Delta \nu_{short}}} {\frac{ \int_{long}^{} B_\nu (T_d) Kabs_\nu d\nu} {\Delta \nu_{long}}}
    \end{equation}
    where the relevant dust properties are captured in the absorption cross-section $Kabs_\nu$ and the integration is carried out over the filter bandwidths $\Delta \nu$. Values for $Kabs_\nu$ were taken from the extinction curve for the Milky Way with $R_V$ = 3.1 of Draine (2003, arXiv:astro-ph/0304488). 
    \item Use $T_d$ and the blurred short wavelength map to calculate the dust optical depth $\tau_\nu$ from:
    \begin{equation}
        F_\nu = \tau_\nu \times B_\nu (\nu_{short}, T_d) 
    \end{equation}
\end{enumerate}

\begin{figure}
     \includegraphics[width=\columnwidth]{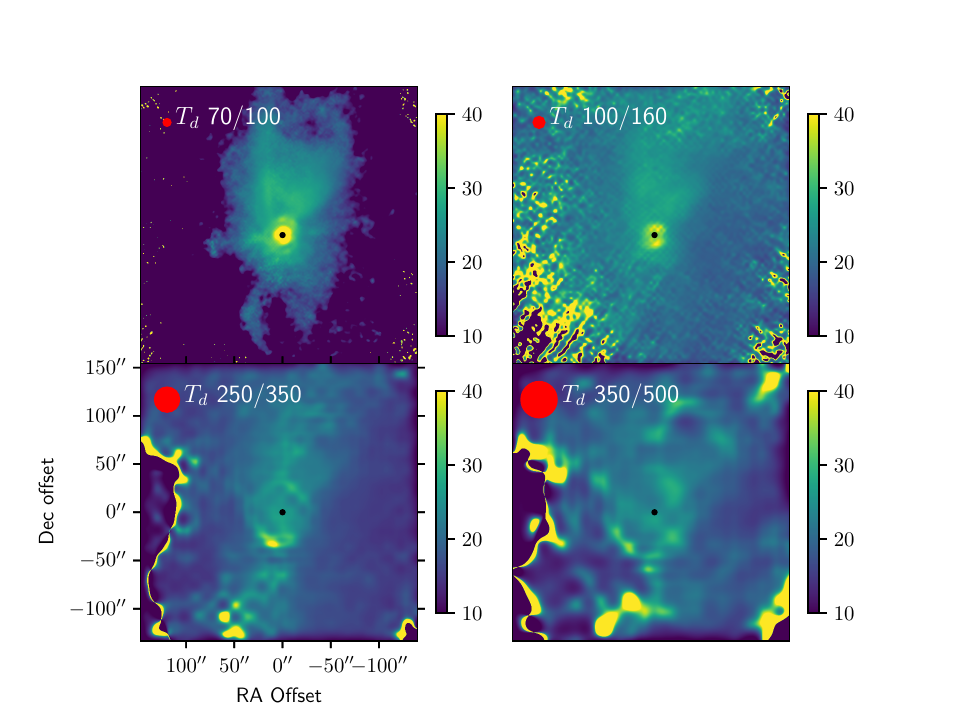}
     \caption{The colour temperature [K] of the far-IR emission at different wavelengths}
     \label{fig:Td}
\end{figure}

The $T_d$ maps thus obtained are shown in Fig.~\ref{fig:Td}.
`Ring' features near R~Mon are dubious and probably caused by failure of the Gaussian beam approximation around this bright source. There is the expected tendency for shorter/longer wavelengths to pick up higher/lower temperature material. The reflection fan has $T_d \sim 30$~K at all wavelengths, standing out from the surroundings where the temperature drops to 15 or 20~K. The faint wisp to the S is another island of warmth.

\begin{figure}
     \includegraphics[width=\columnwidth]{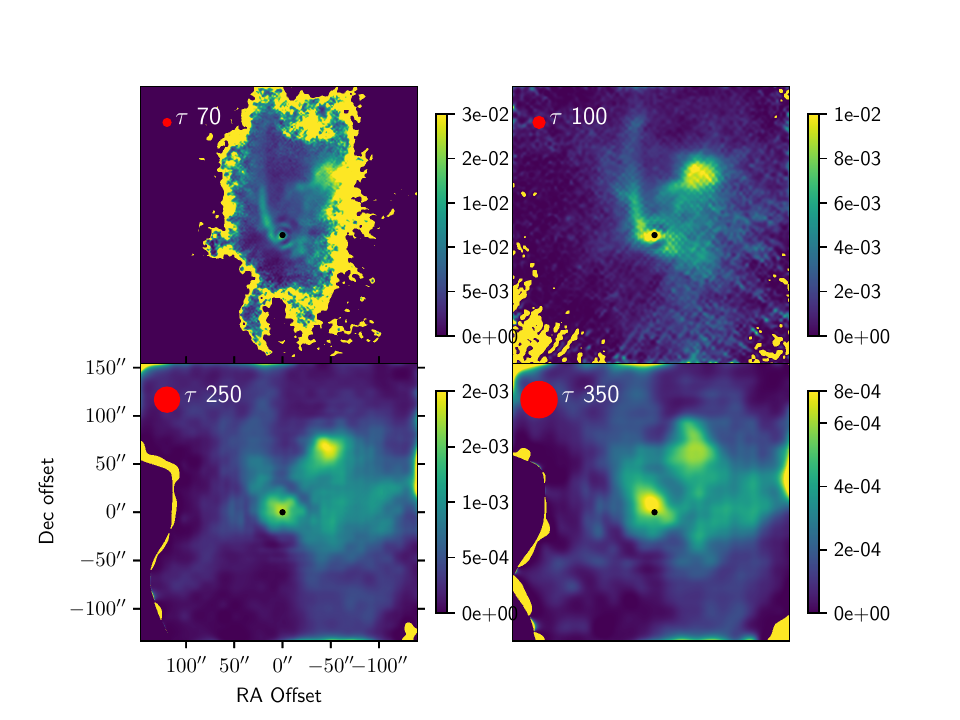}
     \caption{The optical depth of the far-IR emission at different wavelengths}
     \label{fig:tau}
\end{figure}

Corresponding optical depth maps are shown in Fig.~\ref{fig:tau}. Once again the longer wavelength maps are sensitive to lower temperature material. Notable features are a blob at R~Mon, the ridge along the E limb of the fan, the cool blob to the NW, and at longer wavelengths a broad cloud to the W. In general, the optical depth of the material in the fan is 3 times or more greater than in the counterfan to the S. At all wavelengths the faint wisp S of R~Mon is the most prominent feature in the counterfan.

\begin{figure}
    \includegraphics[width=\columnwidth]{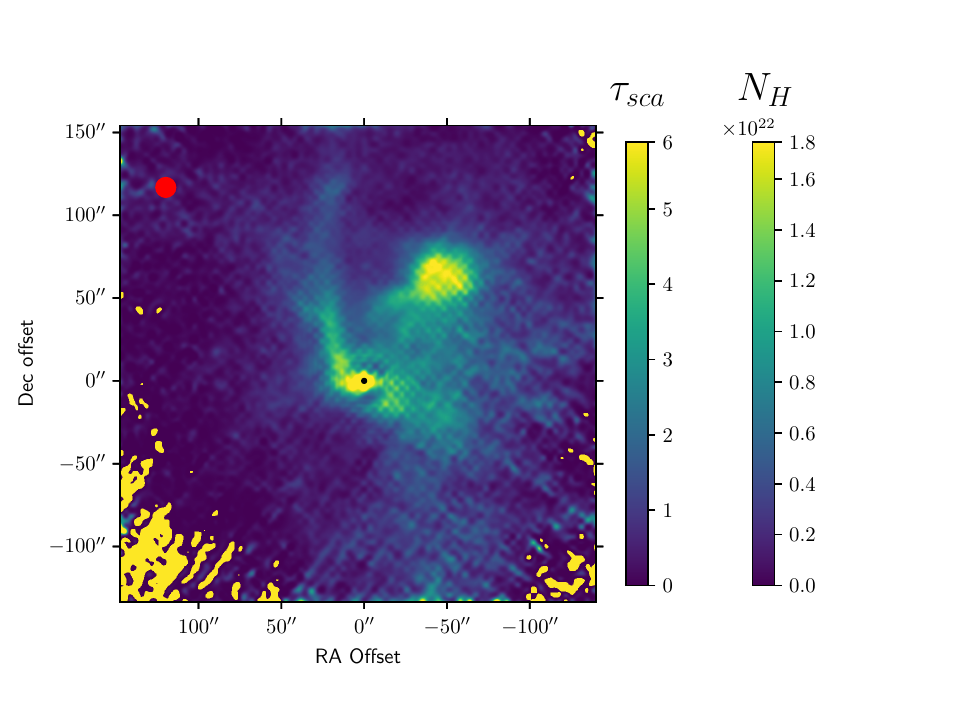}
    \caption{$\tau_{sca}$ at V and column density $N_H$ [cm$^{-2}$] derived from the $100~\micron$ optical depth $\tau_{100}$}
    \label{fig-V-sca-N-H}
\end{figure}

The dust model was combined with the $100~\micron$ optical depth map to calculate the scattering optical depth $\tau_{sca}$ at V, shown by the first colour scale in Fig.~\ref{fig-V-sca-N-H}, using the relation:
\begin{equation}
\tau_{sca} = \tau_{abs} \times albedo / (1 - albedo)
\end{equation}
Scattering by dust is unimportant at far-IR wavelengths but at V~(0.5~nm) the dust albedo rises to 0.68 giving $\tau_{sca}$ ($\sim 2 \times \tau_{abs}$) a value of 1 or 1.5 across the fan so that it appears bright in {\em scattered} starlight from R~Mon.

The gas column density is related to $\tau$ by:
\begin{equation}
    \label{eq:column}
    N_H = \frac{\tau_\nu} {Kabs_\nu \times m\_dust\_per\_H}
\end{equation}
where $Kabs_\nu$ is the absorption cross section per mass of dust and $m\_dust\_per\_H = 1.398\times10^{-26}g$ for the dust model used. The $N_H$ column density derived from the $100~\micron$ optical depth is shown by the second color scale in Fig.~\ref{fig-V-sca-N-H}. At a point $30\arcsec$ N of R~Mon the column density $\sim5\times10^{21}$~cm$^{-2}$ which, combined with the maximum screen thickness from the ripple analysis ($3\times10^{16}$~cm), implies $n_H > 1.7\times10^5$~cm$^{-3}$.

The cool blob $80\arcsec$ to the NW of R~Mon in all IR bands has no obvious visible counterpart but its CO line velocity indicates it is associated with the nebula (Sec.~\ref{sec:molecular}). It coincides with a very red area in images at visible wavelengths, so it is possibly just a massive section of undisturbed material in front of the fan.

\subsection{The molecular nebula}
\label{sec:molecular}
Early CO~(J=1-0) maps of the object by \citet{canto1981carbon} placed it in a small elongated molecular cloud $\sim0.8\times0.4$~pc in size, with the reflection nebula marking the wall of a hollow cone carved out by winds from R~Mon. More recent observations by \citet{Sandell_2020} use optically thin C$^{18}$O~(2-1) emission to measure the rest cloud velocity as $V_{\mathrm{LSR}}=9.5$~km~s$^{-1}$, and their high spatial resolution maps in CO and $^{13}$CO~(3-2) confirm cavities in emission at this velocity to N and S of the star, with that to the N matching the extent of the reflection fan. A ridge of emission curving NE from the star along the fan edge matches the bright limb seen at visible and far-IR wavelengths, where the screen curves into the line of sight. They estimate the total mass of the cloud to be $\sim70M_\odot$. The CO velocity varies smoothly over the fan and is blue-shifted by up to a few km~s$^{-1}$ relative to the rest cloud. A broader CO peak to the NW coincides with the cool blob seen in the far-IR. 

The correspondence between reflection fan and blue-shifted CO suggests that gas in the screen is responsible for the line flux. Lack of emission from the fan centre at the cloud rest velocity would be consistent with the ambient material there having been swept aside by the outflow into a compressed shell that is moving relative to the system.

Is there enough CO in the screen to produce the observed lines? We model it as a dense slab ($n_H = 1.7\times10^5$~cm$^{-3}$) filling the telescope beam, with abundances
of $\mathrm{CO/H_2}=10^{-4}$ and $\mathrm{^{12}C/^{13}C} =50$. The density exceeds the critical density of CO for rotational levels up to J=4 so that the level populations are near LTE and the line brightness is just a function of the column density and excitation temperature T$_{ex}$.

Within the uncertainty due to comparing observations from different telescopes, the available $^{12}$CO line measurements are consistent with the model slab having $T_{ex}\sim20$~K, as found by \citet{Sandell_2020}. However, all the model lines are optically thick and of similar brightness at $8-10$K, and the $^{13}$CO observations do not fit this picture; \citet{bachiller} measure  $^{13}$CO~(1-0) to be $\sim 1$ K.

For model to match observation the abundance of $^{13}$CO has to be depleted by a factor of 30 to make the problem line optically thin. This is possible given that the volume is a photon-dominated region exposed to the interstellar UV field and open to isotope selective photodestruction \citep{1982ApJ...255..143B}; for example, \citet{Ripple2013COAV} found $^{13}$CO to be depleted by a factor of up to 20 in translucent ($A_V<3$~mag) parts of the Orion~B cloud. Alternatively, the model premise that the slab has uniform $T_{ex}$ could be wrong, with the CO and $^{13}$CO line emission being dominated by a warm cloud surface and cold core respectively.

We conclude that the blue-shifted CO emission over the reflection fan could be produced by gas in the screen though the details are unclear.

\section{Where is the counterfan?}
At visible wavelengths NGC~2261 is a cometary nebula with R~Mon at its head and the fan just the brightest part of a long tail to the N. At longer wavelengths the system is obviously bipolar, with the star at the centre and extensions to both N and S. Why is the counterfan so dim in visible light?

The {\it Herschel} observations rule out the possibility that the counterfan is obscured by foreground cloud material. Such would surely appear in the far-IR maps but Fig.\ref{fig:tau} shows $\tau_{250}\sim0.0001$ in the area, corresponding to $\tau_V\sim 0.2$, insufficient to hide bright emission. In addition, an obscured counterfan would appear as bright as the fan at $70\micron$ in Fig.~\ref{fig:farIR}, but it does not.

\begin{figure}
    \includegraphics[width=\columnwidth]{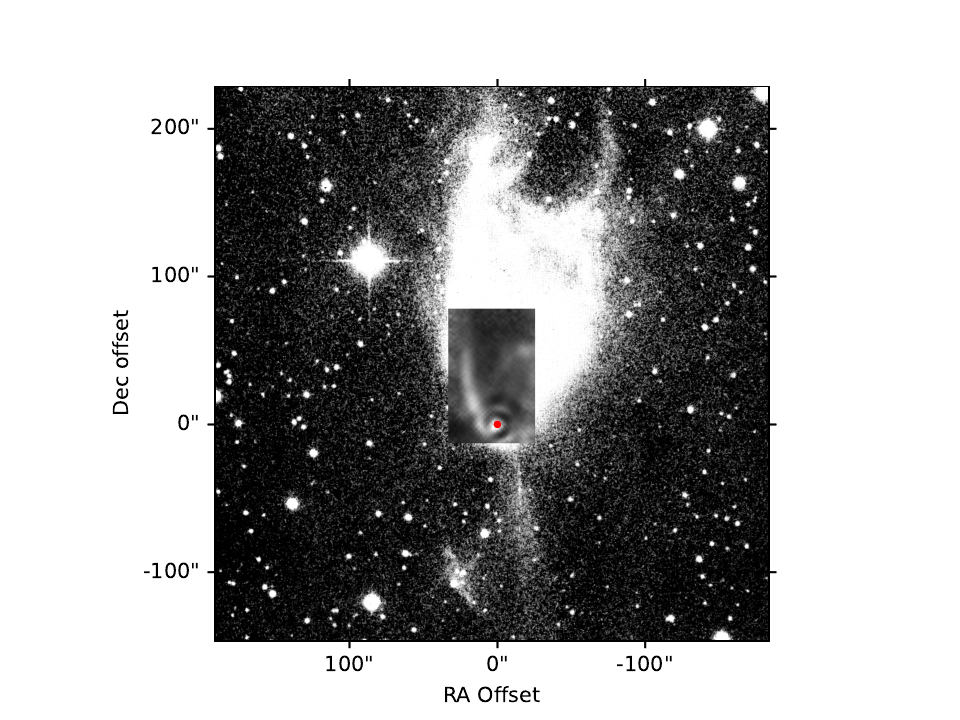}
    \caption{Image composite made up of a deep B photograph of the nebula from \citet{10.1093/mnras/217.1.31} with an insert showing $\tau_{70}$ over R~Mon and the visible fan. A red dot shows the position of the star.}
    \label{fig-aat-herschel}
\end{figure}

The answer probably lies in several factors working together. The first is illustrated in Fig.~\ref{fig-aat-herschel}, which shows that the visible counterfan is present but very faint. The far-IR ridge marking the boundary of the N cavity is mirrored, relative to R~Mon, by a faint visible wisp that could be the edge of a matching S cavity. However, this feature is not seen in the far-IR and generally the optical depth maps in Fig.\ref{fig:tau} show roughly 3 times less material in the counterfan area than the fan. The evidence is that there is simply not as much material to illuminate in the S.

Second, it is clear that light from R~Mon is blocked in most directions by material close to the star. For example, the W limb of the fan looks like a shadow edge. That starlight to the S is blocked close to R~Mon today is proven by the fact that it has not always been so. Photographs taken between 1916 and 1948 show a bright `anti-tail' projecting up to $5\arcsec$ S from R~Mon \citep{1966AJ.....71..224J}. This had faded to invisibility by the end of the period while the fan brightened, consistent with slow relative motion between the clouds and illuminating star.

Third, the CO outflow N of R~Mon is blue-shifted while that to the S is red-shifted. Were the distribution of material mirrored about the star as this suggests, then the N fan would shine by starlight forward scattered by dust and the counterfan by backward scattered. At visible wavelengths, the scattering efficiency falls by a factor of 8 between forward ($\phi\lesssim30^{\circ}$) and backward ($\phi\gtrsim90^{\circ}$) \citep{draine2003scattering}. Other things being equal, this would make the counterfan significantly fainter than the fan.

\section{Conclusions}
A 7 year long monitoring campaign of NGC~2261 shows it to be crossed by frequent light ripples emanating from R~Mon. These are distinct from the occasional large shadow variations for which the object is famous. The ripple behaviour is consistent with variable light output from R~Mon travelling outward and reflecting off a thin screen stretching up from the star. 

Detailed modelling indicates that the screen is a roughly parabolic hollow cone with R~Mon at the vertex, of which only the near and E sides are illuminated. The E limb of the nebula marks the cone edge as it curves into the line of sight, whereas the sharp and straight W limb is a shadow cast across the screen by something close to the star. The shape of the screen is such that shadows will be cast on it by objects crossing R~Mon in a latitude band $\sim15^{\circ}$ wide.

Herschel maps of far-IR thermal emission from dust and radio observations of CO rotational lines are consistent with the screen being an expanding shell driven by the outflow from R~Mon. They show that the absence of a counter-fan to the S of R~Mon is due to a lack of material to illuminate and blocking of light near R~Mon, rather than foreground obscuration.

The nebula's large shadow variations and the changes in R~Mon's brightness driving the ripples on the fan will be the subjects of a future paper. 

This study shows the utility of long-term monitoring of this class of object and the type of analysis it makes possible. Such observation has been difficult before now, but with the arrival of the ZTF and the promise of the Vera Rubin Observatory \citep{2019ApJ...873..111I} the future looks bright.

\section*{Acknowledgements}

JFL thanks Bill Glencross and Derek McNally (UCL) for the loan of a copy of the Hall movie, Dick Jennings (UCL) for his introduction to far-IR astronomy, and Paul Slade for his friendship lang syne. We thank the referee, John Bally, for various questions and suggestions that let to improvements in the paper.

This work makes use of observations from the Las Cumbres Observatory global telescope network. We thank the National Schools Observatory, Faulkes Telescope Project, and Tom Polakis for access to their data. 

The Liverpool Telescope is operated on the island of La Palma by Liverpool John Moores University in the Spanish Observatorio del Roque de los Muchachos of the Instituto de Astrofisica de Canarias with financial support from the UK Science and Technology Facilities Council.

{\it Herschel} is an ESA space observatory with science instruments provided by European-led Principal Investigator consortia and with important participation from NASA.

\section*{Data Availability}

The LCOGT and Faulkes Telescope data used can be found in the Science Archive of the Las Cumbres Observatory (https://archive.lco.global). LT data are in the Liverpool Telescope Data Archive (https://https://telescope.livjm.ac.uk/). {\em Herschel} data are available at the Herschel Science Archive (https://archives.esac.esa.int/hsa/whsa/).



\bibliographystyle{mnras}
\bibliography{mnras_2261}









\bsp	
\label{lastpage}
\end{document}